\documentclass[twocolumn,english]{revtex4-2}
\usepackage[T1]{fontenc}
\usepackage{textcomp}
\usepackage[latin9]{luainputenc}
\usepackage{color}
\usepackage{babel}
\usepackage{amstext}
\usepackage{graphicx}
\usepackage[pdfusetitle,
 bookmarks=true,bookmarksnumbered=false,bookmarksopen=false,
 breaklinks=false,pdfborder={0 0 1},backref=false,colorlinks=false]
 {hyperref}

\makeatletter

\providecommand{\tabularnewline}{\\}

\makeatother

\begin{document}
\title{Long-range exchange coupling in a magnetic multilayer system}
\author{L. O. Souza }
\email{leandro.souza@fisica.ufmt.br}

\affiliation{Instituto de Física - Universidade Federal de Mato Grosso, 78060-900,
Cuiabá, Mato Grosso, Brazil.}
\author{R. A. Dumer}
\email{rafaeldumer@fisica.ufmt.br}

\affiliation{Instituto de Física - Universidade Federal de Mato Grosso, 78060-900,
Cuiabá, Mato Grosso, Brazil.}
\author{M. Godoy}
\email{mgodoy@fisica.ufmt.br}

\affiliation{Instituto de Física - Universidade Federal de Mato Grosso, 78060-900,
Cuiabá, Mato Grosso, Brazil.}
\begin{abstract}
In this study, we explored the magnetic coupling in a multilayer system
consisting of thin layers separated by a distance $d$. We have employed
Monte Carlo (MC) simulations to calculate the thermodynamic quantities
such as the magnetization per spin $m_{L}^{\mu}$, magnetic susceptibility
$\chi_{L}^{\mu}$, and the reduced fourth-order Binder cumulant $U_{L}^{\mu}$
as a function of temperature $T$ and for several values of lattice
size $L$. These quantities were obtained for each layer ($\mu=l)$,
for the bulk system ($\mu=b)$, as a function of interlayer distance
$d$ and the parameter $\alpha$ that defines the scale length of
the exponential decay of the interaction. Furthermore, we applied
the finite-size scaling theory to calculate the critical exponents.
Our results reveal that the system exhibits 2D Ising exponents when
the layers are sufficiently separated. On the other hand, in the compact
limit, where $d$ equals the distance between two adjacent sites within
the same layer, our bulk results show that the system exhibits 3D
Ising critical exponents, provided that the number of layers $L_{z}$
increases proportionally to the layer sizes. Even with the separation
increasing up to $d=2.0$, the layers are so correlated that the set
of critical exponents retains the values of 3D critical exponents.
However, when the number of layers $L_{z}$ remains fixed, even in
the compact limit with periodic boundary conditions, only the exponent
$\beta$ aligns closely with the predicted literature values, on the
other hand, the other exponents show significant deviations.
\end{abstract}
\maketitle

\section{Introduction}

Over the past thirty years, metallic spintronics has established itself
as a significant field, with one of the most notable applications
being the use of the Giant Magnetoresistance (GMR) effect in magnetic
metallic multilayers for data storage \citep{1,2}. These materials
feature ferromagnetic layers whose exchange coupling between adjacent
layers oscillates in sign as the thickness of the non-magnetic spacer
increases \citep{3}. The next generation of spintronic materials
aims to manipulate spin degrees of freedom at lower carrier densities
and achieve control through applied electric voltages, with semiconductor-based
materials expected to play a pivotal role, see, e.g., Refs. \citep{4}
and \citep{5}.

In this context, diluted magnetic semiconductors (DMS) have emerged
as promising materials for generating spin-polarized carriers that
can be manipulated for injection \citep{6}. Among these, (Ga, Mn)As
stands out as one of the most well-studied DMS. When a Mn atom substitutes
a Ga atom, it introduces a localized magnetic moment along with one
hole \citep{6,7}. However, many of these holes are compensated by
defects. The remaining uncompensated holes interact antiferromagnetically
with each Mn local moment, leading to ferromagnetic order below a
critical temperature, $T_{c}(x)$, which varies non-monotonically
with the concentration $x$ of Mn \citep{8}. 

Advances in experimental techniques have significantly increased the
observed maximum critical temperature, from $60K$ \citep{9} to $110K$
\citep{8}, and further improvements in post-growth annealing processes
have pushed $T_{c}$ to $184K$ \citep{10}. Additionally, developments
in deposition techniques have enabled the growth of digital ferromagnetic
hetero-structures (DFH\textquoteright s), where submonolayer planes
of MnAs are inserted into GaAs layers using molecular beam epitaxy
\citep{11}. Since holes can potentially correlate the magnetization
in adjacent MnAs layers, a systematic study of interlayer coupling
is essential.

With this in mind, we intend to investigate the interplay between
the layering and ordering of localized magnetic moments placed regularly
on equally spaced layers, coupled through hole-mediated exchange coupling.
The broad features of this exchange coupling can be determined by
following Refs. \citep{12,13,14}. The long-range coupling constant
of the form $J_{ij}=J_{0}\exp(-2r_{ij}/a_{B})$ \citep{12}, i.e.,
decaying algebraically with the interspin distance, that favors ferromagnetic
ordering at low temperature. Here, $J_{0}$ defines the exchange energy
scale, and $a_{B}$ is the effective Bohr radius, for instance, with
typical values for Mn in bulk GaAs being $J_{0}=15meV$ and $a_{B}=7.8\text{Å}$
\citep{12}. Although this impurity band perspective may not be entirely
suitable for low carrier densities, it nonetheless reflects the localized
character of holes in (Ga,Mn)As \citep{15}. Thus, when holes are
integrated out, the interaction between pairs of localized spins primarily
exhibits an exponential decay relative to their separation distance.

For magnetic models involving Heisenberg spins with significant quantum
numbers (such as $S=5/2$ for Mn spins, which considerably affect
the magnetic response), the quantum corrections are typically minor.
Therefore, we can use a classical vector spin model. In this context,
considering the Ising spins and utilizing MC simulations is a reasonable
approach \citep{16,17}.

This paper is structured as follows. Section \ref{sec:Model}, we
detail the model used and outlines the MC simulation methods. Section
\ref{sec:Results}, we present and discuss the results. Finally, in
Section \ref{sec:Conclusions}, we display our conclusion with a summary
of our findings.

\section{Model and Methodology\protect\label{sec:Model}}

In line with our previous discussion, the classical spin Hamiltonian
that we have studied is written by the following form:

\begin{equation}
\mathcal{H}=-\sum_{i,j}J(r_{ij})S_{i}S_{j},\label{eq:1}
\end{equation}
where $S_{i}=\pm1$ is the Ising spin variable at the position $R_{i}$
of the lattice. While an oscillatory (i.e. alternating in sign) behavior
of $J(r_{ij})$ has been ruled out by first-principles calculations
\citep{18,19}. Therefore, it is instructive to investigate the effects
of an additional power-law decay, which is reminiscent of the RKKY
interaction \citep{20,21,22}. That is, we propose a plausible configuration
for the magnetic coupling $J(r_{ij})$:

\begin{equation}
J(r_{ij})=\frac{J_{0}}{r_{ij}^{3}}\exp\left(-\frac{r_{ij}}{\alpha}\right),\label{eq:2}
\end{equation}
where $r_{ij}$ is the distance between the spins at site $i$ and
$j$ , and $J(r_{ij})$ is the interaction strength. Here, the distance
$r_{ij}$ and the parameter $\alpha$ are expressed in units of the
underlying lattice spacing, which, for our purposes here, it suffices
to consider as being of the order of the effective Bohr radius $a_{B}$
mentioned above. The sum in Eq. (\ref{eq:1}) is performed over all
pairs spins. These atoms reside on $L_{z}$ square lattice layers,
each of which with $N=L\times L$ sites, and which are separated from
each other by non-magnetic spacers of thickness $d$, as illustrated
in Fig. \ref{fig:1}.
\begin{center}
\begin{figure}
\begin{centering}
\includegraphics[scale=0.14]{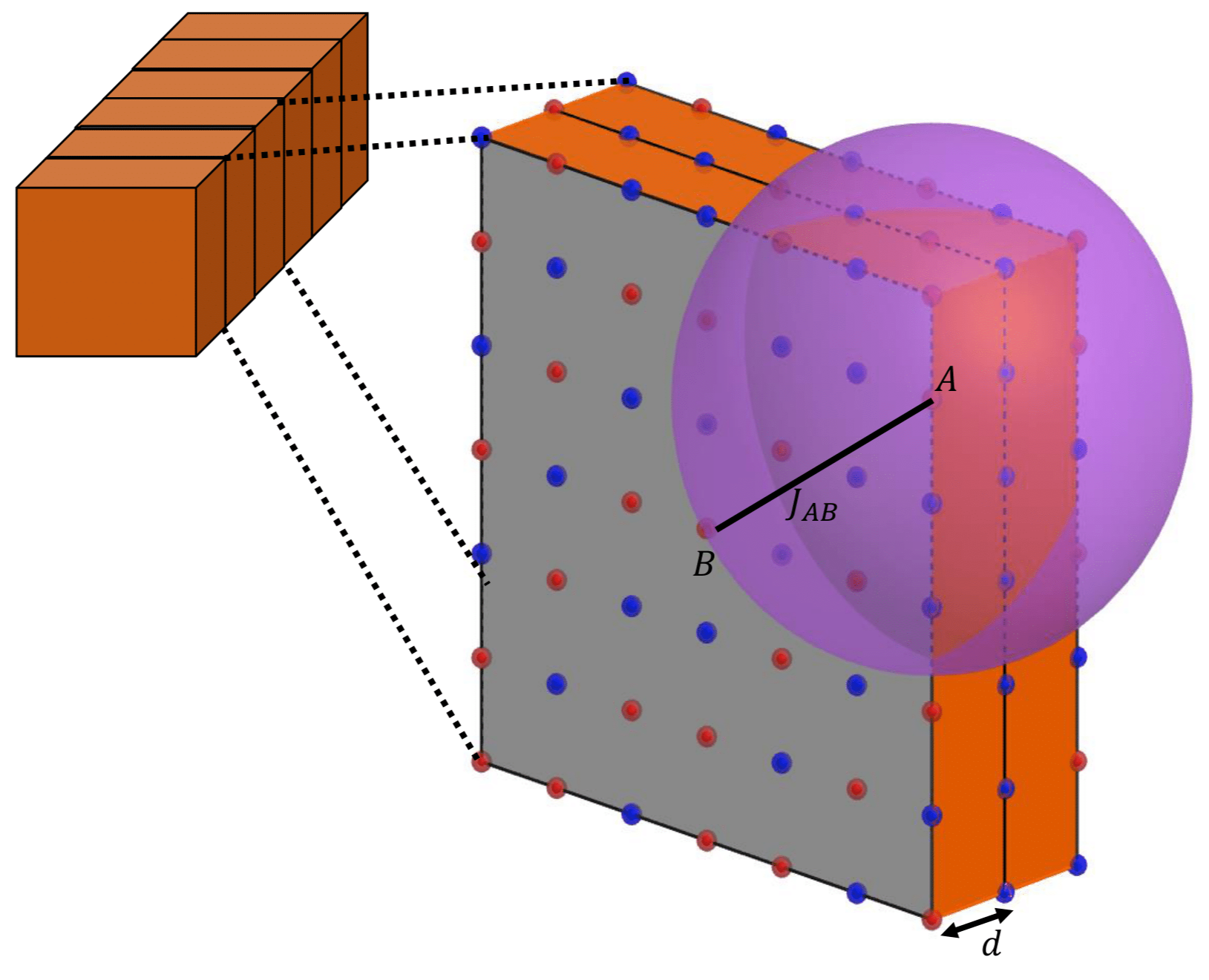}
\par\end{centering}
\caption{Schematic representation of a two-dimensional multilayers system with
thickness $d$. $J_{AB}$ is the long-range magnetic coupling $J(r_{ij})$
defined in Eq. \ref{eq:2}. \protect\label{fig:1}}

\end{figure}
\par\end{center}

The simulations, especially for large lattices, are extremely computationally
costly, given that the interaction $J(r_{ij})$, as defined by Eq.(\ref{eq:2}),
occurs among all sites of the lattice. To overcome this limitation,
we have set a cutoff in $J(r_{ij}^{\textrm{cutoff}})$ when the exponential
decay has rendered the interactions sufficiently tenuous to be disregarded.
Therefore, the interaction occurs within a cluster of spins smaller
than the entire lattice. Let $N_{is}(R)$ be the number of new interactions
that can be added at a site within an interaction radius $R$. Despite
$N_{is}$ increasing with the interaction radius $R$, as shown in
Fig. \ref{fig:2}(a), the value of $J(R)$ decays exponentially. For
$R>2.5$, the value of $J(R)$ falls below $3\%$, as demonstrated
in Fig. \ref{fig:2}(b). When considering the contribution of all
interacting sites, as illustrated in Fig. \ref{fig:2}(c), we can
observe that setting a maximum radius as $R=2.5$ is reasonable, since
the most relevant interactions are among $94\%$ of the entire lattice
interaction. Thus, we set a cutoff in $J(r_{ij})$ for $r_{ij}^{\textrm{cutoff}}=2.5$,
i. e., in other words we have used $J(r_{ij}\geq2.5)=0$.\textcolor{red}{{} }
\begin{center}
\begin{figure*}
\begin{centering}
\includegraphics[scale=0.4]{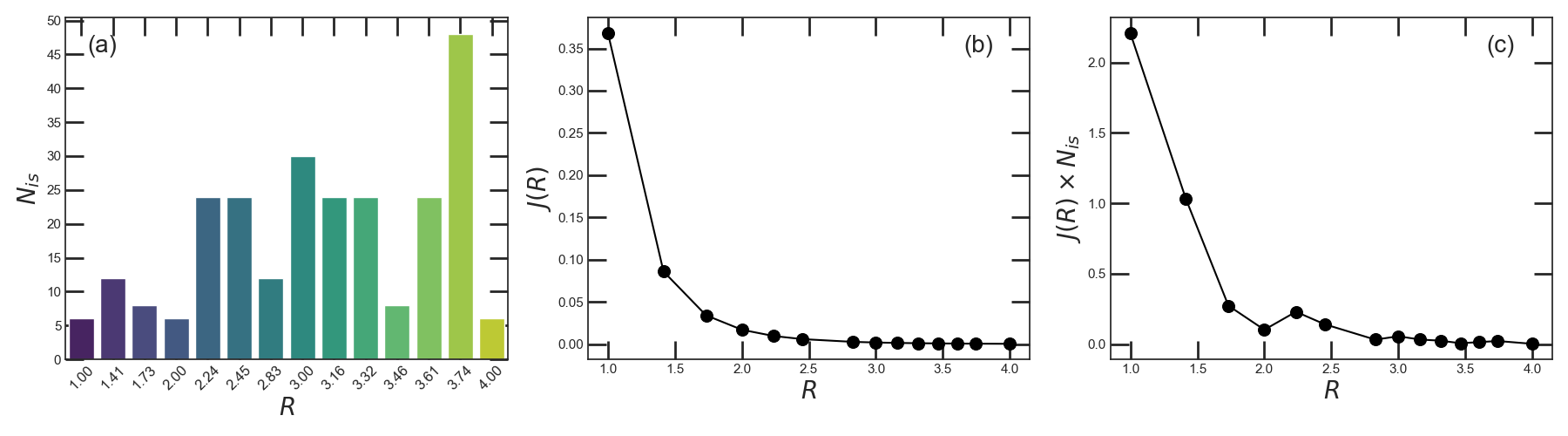}
\par\end{centering}
\caption{(a) Number of new interactions $N_{is}$ as a function of radius $R$.
(b) Magnetic couplings $J(R)$ as a function of the radius $R$. (c)
Total contribution of the magnetic coupling $J(R)\times N_{is}$ for
spins at a distance $R$. The plots (a), (b), and (c) were obtained
for $\alpha=1.0$ and $d=1.0$. \protect\label{fig:2}}
\end{figure*}
\par\end{center}

Since in the limit $d\rightarrow\infty$, the system presents a collection
of independent planes. Therefore, in our MC simulations, we have calculated
the thermodynamic quantities for each layer and the bulk system. Consequently,
we can define the following parameter:

\begin{equation}
M_{\mu}=\sum_{i=1}^{N_{\mu}}S_{i},\label{eq:3}
\end{equation}
such that when the subscript $\mu=l$ and $\mu=b$ represent the layer
and the bulk, respectively. For $\mu=l$, the sum runs over the $N_{l}=L^{2}$
sites within the layer and for $\mu=b$, the sum runs over the $N_{b}=N_{l}L_{z}=L^{2}L_{z}$
sites within the bulk. We also have considered the kth-moment of the
magnetization as

\begin{equation}
\left(m_{L}^{\mu}\right)^{k}\equiv\frac{1}{N_{\mu}^{k}}M_{\mu}^{k}.\label{eq:4}
\end{equation}
The first-moment yields the magnetization per site,

\begin{equation}
m_{L}^{\mu}=\frac{1}{N_{\mu}}\left\langle \sum_{i=1}^{N_{\mu}}S_{i}\right\rangle ,\label{eq:5}
\end{equation}
magnetic susceptibility is given by

\begin{equation}
\chi_{L}^{\mu}=\frac{N_{\mu}}{k_{B}T}\left[\left\langle \left(m_{L}^{\mu}\right)^{2}\right\rangle -\left\langle m_{L}^{\mu}\right\rangle ^{2}\right],\label{eq:6}
\end{equation}

In ordetr to obtain the critical temperatures we can also use the
fourth-order Binder cumulant \citep{17}:

\begin{equation}
U_{L}^{\mu}=1-\frac{\langle m_{\mu}^{4}\rangle}{3\langle m_{\mu}^{2}\rangle^{2}}\label{eq:7}
\end{equation}
where $\left\langle \dots\right\rangle $ denotes the thermal average
over the MC simulation.

The magnetic layers are separated by non-magnetic spacers of thickness
$d$, as illustrated in Fig. \ref{fig:1}. We vary $d$ and perform
systematic MC simulations to calculate the thermodynamic quantities
of interest (Fig. \ref{fig:3}), and the critical exponents for the
layers and the bulk.

Near the critical temperature $T_{c}$, the previously defined quantities
conform to the finite-size scaling relations as follows:

\begin{equation}
m_{L}^{\mu}=L^{-\beta/\nu}m_{0}(L^{1/\nu}\epsilon),\label{eq:9}
\end{equation}

\begin{equation}
\chi_{L}^{\mu}=L^{\gamma/\nu}\chi_{0}(L^{1/\nu}\epsilon),\label{eq:10}
\end{equation}

\begin{equation}
U_{L}^{\mu\prime}=L^{1/\nu}U_{0}(L^{1/\nu}\epsilon)/T_{c},\label{eq:11}
\end{equation}
where $\epsilon=(T-T_{c})/T_{c}$. In the quations above $m_{0}(L^{1/\nu}\epsilon)$,
$\chi_{0}(L^{1/\nu}\epsilon)$, and $U_{0}(L^{1/\nu}\epsilon)$ are
the scaling functions, and $\beta$, $\gamma$, and $\nu$ are the
critical exponents for magnetization, magnetic susceptibility, and
length correlation, respectively, and $U_{L}^{\mu\prime}$ is the
derivative of $U_{L}^{\mu}$ with relation to $T$.
\begin{center}
\begin{figure*}
\begin{centering}
\includegraphics[scale=0.4]{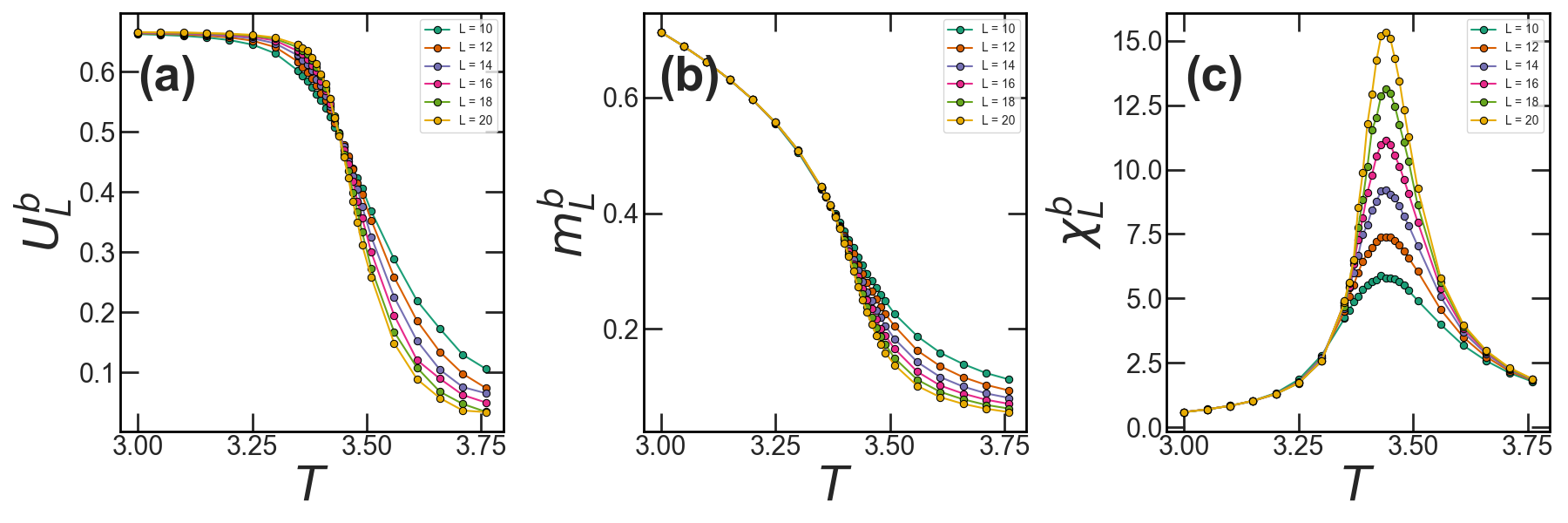}
\par\end{centering}
\caption{(a) Fourth-order Binder cumulant $U_{L}^{b}$ , (b) magnetization
$m_{L}^{b}$, and (c) susceptibility $\chi_{L}^{b}$ versus temperature
$T$ for several lattice sizes $L$, as indicated in the figures.
The results were obtained for $\alpha=1.0$, $d=1.0$, $L_{z}=10$,
$N_{b}=10L^{2}$. The critical temperature obtained by intersection
of $U_{L}^{b}$ (a) is given by $T_{c}=3.44\pm0.02$. The error bars
are smaller than the symbol size. \protect\label{fig:3}}
\end{figure*}
\par\end{center}

\section{Results and Discussion \protect\label{sec:Results}}

In the following, we have considered layers with linear sizes ranging
from $10\leq L\leq20$, while the stack size is fixed at $L_{z}=10$.
Additionally, we examined the cases where $10\leq L_{z}\leq20$ with
$L_{z}=L$, $N_{l}=L^{2}$, and $N_{b}=L_{z}N_{l}=LL^{2}=L^{3}$.
We have allowed from $10^{5}$ to $10^{6}$MC steps/spin to reach
equilibrium, and averages have been obtained over $10^{6}$ to $10^{7}$
steps/spin. Fig. \ref{fig:3} illustrates the temperature dependence
of some bulk thermodynamic quantities, with the parameter $\alpha=1.0$
and the interlayer distance $d=1.0$, for several lateral sizes $L$
and $L_{z}=10$. In Fig. \ref{fig:3}(a), we display the Binder cumulant,
where we can observe that this quantity tends to 2/3 deep in the ordered
phase of the system and decreases to zero well into the disordered
phase. It can be shown that, for sufficiently large system sizes,
the Binder cumulants are rather insensitive to the system size, and
their curves cross each other at the same point $T_{c}$, regardless
of $L$, thus providing an estimate of the critical point in the thermodynamic
limit ($L\rightarrow\infty$). Therefore, as determined by the common
intercept of the curves for different system sizes we have found $T_{c}=3.44\pm0.02$.
This estimate is consistent with those obtained from the magnetization
data, as well as with those signaled by the maxima in the susceptibility;
see Fig. \ref{fig:3}(b) and (c). For each $L$, the inflection point
in $m_{L}^{b}$ versus $T$ provides the existence of second-order
phase transitions occurring between ordered and disordered phase,
and can be see in Fig. \ref{fig:3}(b). 

In Fig. \ref{fig:4}, we display the temperature dependence of the
layers susceptibility $\chi_{L}^{l}$, for several interlayer separations.
Now, in the compact limit for $d=1.0$, the peak in $\chi_{L}^{l}$
is located at $T_{c}\approx3.44$ and as $d$ increases, the critical
temperature decreases. Notably, once the distance $d$ between the
layers surpasses $2.5$, the layers act independently, halting the
decrease in critical temperature.
\begin{center}
\begin{figure}
\begin{centering}
\includegraphics[scale=0.33]{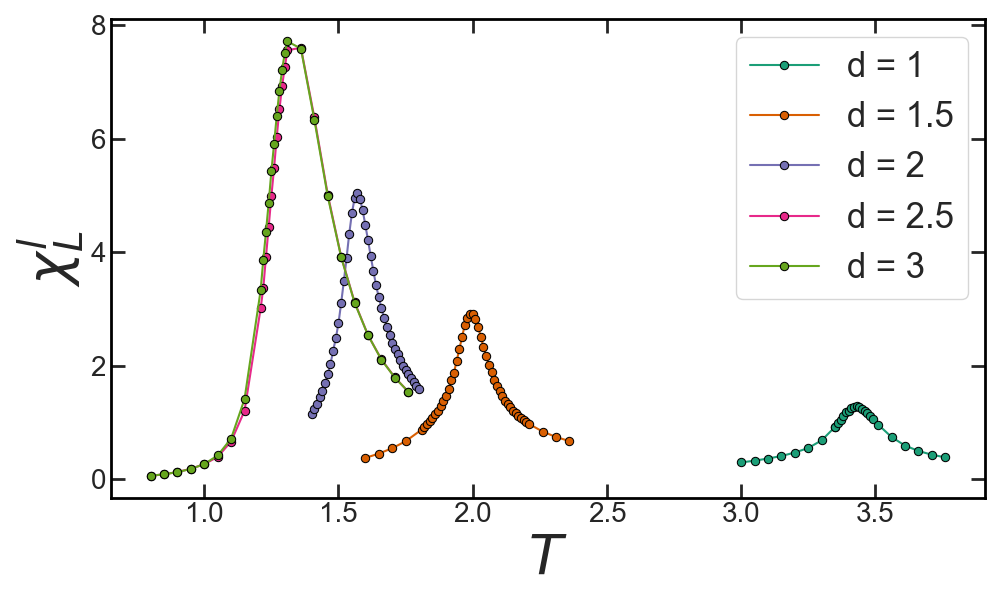}
\par\end{centering}
\caption{Bulk magnetic susceptibility $\chi_{L}^{l}$ versus temperature $T$
for different values of interlayer separations $d$, as indicated
in the figure. The results were obtained for $\alpha=1.0$ and $L=14$.
\protect\label{fig:4}}
\end{figure}
\par\end{center}

We also studied the curves of $m_{L}^{l}$, $U_{L}^{l}$, and $\chi_{L}^{l}$
versus temperature $T$, as shown in Fig. \ref{fig:5}, in the vicinity
of the phase transition point. These quantities were obtained from
the average value of each layer since the isolated layers presented
the same critical behavior.
\begin{center}
\begin{figure*}
\begin{centering}
\includegraphics[scale=0.35]{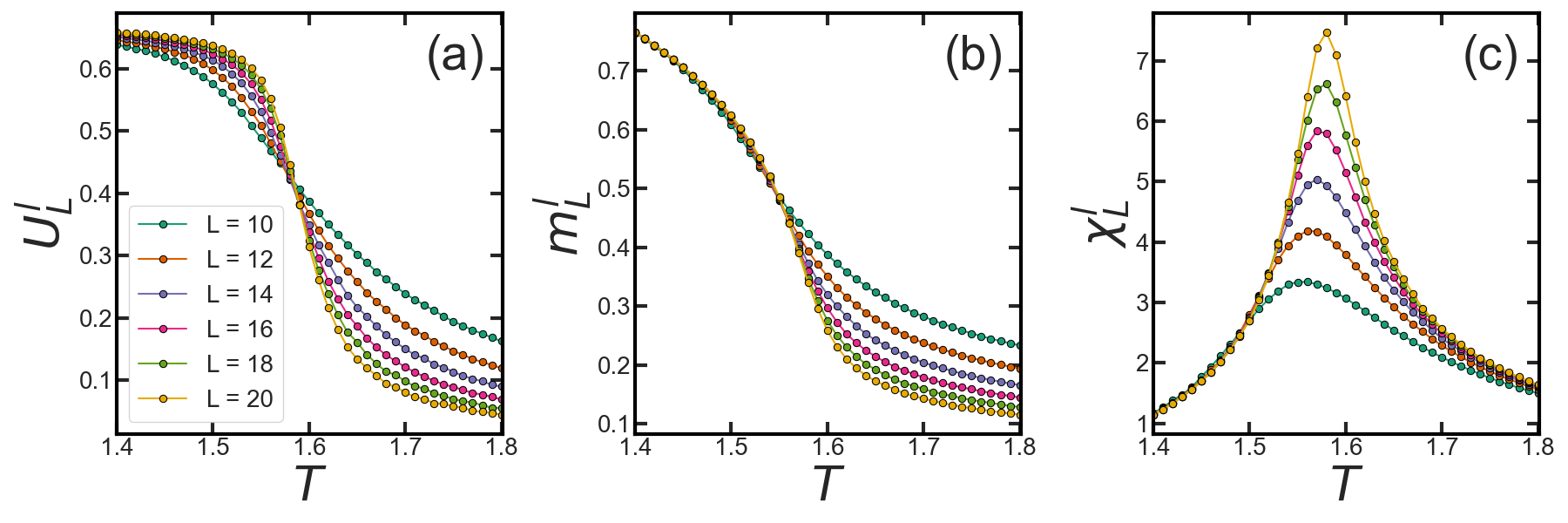}
\par\end{centering}
\caption{(a) Fourth-order Binder cumulant $U_{L}^{l}$, (b) magnetization $m_{L}^{l}$,
and (c) susceptibility $\chi_{L}^{\mu}$ versus temperature $T$ for
several lattice sizes $L$, as indicated in panel (a). The results
were obtained for $\alpha=1.0$, $d=2.0$, and with the exchange coupling
given by Eq. (\ref{eq:2}). The critical temperature obtained of $U_{L}^{b}$
(a) is given by $T_{c}=1.585\pm0.010$. The error bars are smaller
than the symbol size. \protect\label{fig:5}}
\end{figure*}
\par\end{center}

\begin{center}
\begin{figure}
\begin{centering}
\includegraphics[scale=0.3]{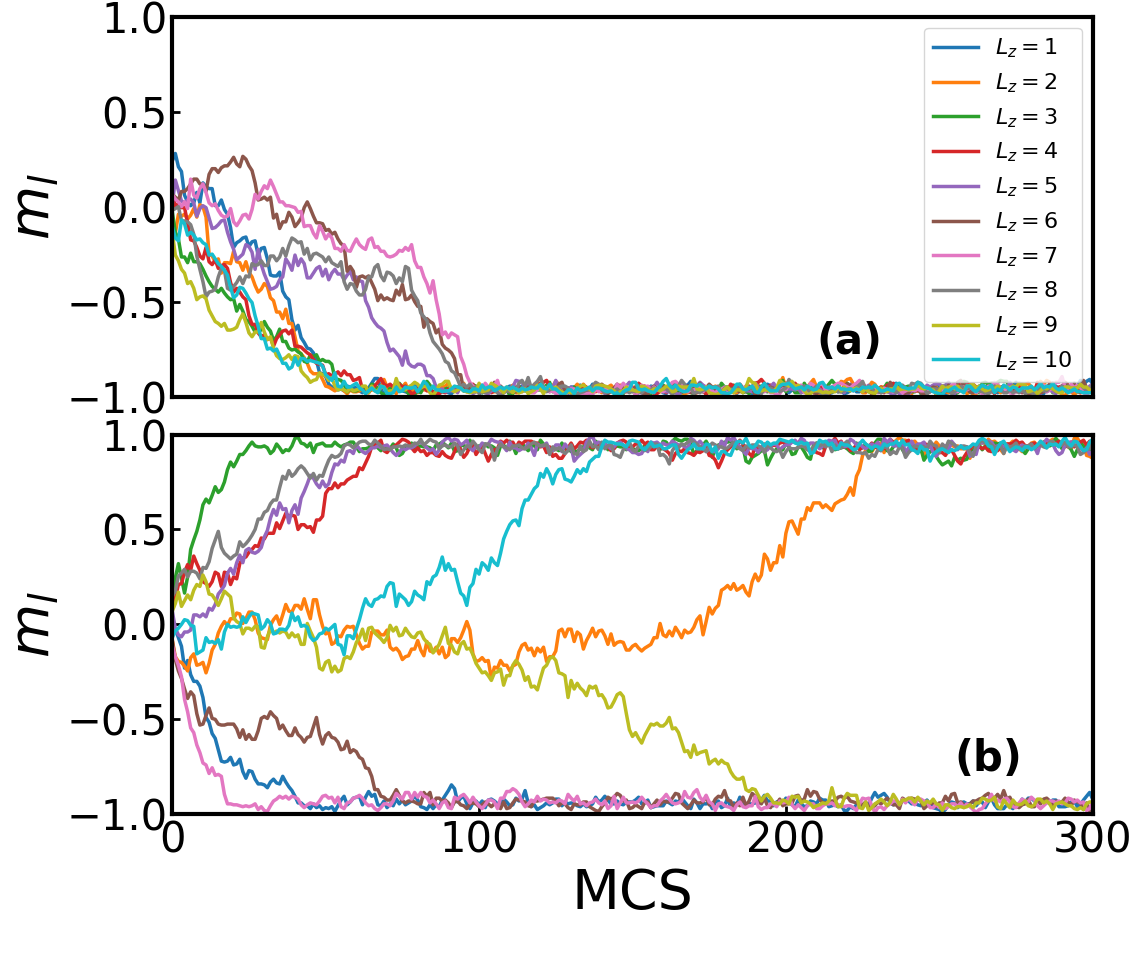}
\par\end{centering}
\caption{Layer magnetization $m_{L}^{l}$ as a function of MCS for several
lattice sizes $L_{z}$, as indicated in panel (a). The results were
obtained for (a) $d=2.0$ and (b) $d=3.0$. \protect\label{fig:6}}
\end{figure}
\par\end{center}

For a finite lattice, the system can transition between positive and
negative magnetization states, and for a statistically large sample,
the resulting magnetization is zero. As a result, magnetization is
calculated using the average of the absolute value of the magnetic
moment. When we consider a multilayer system, if the layers are close
($d<2.5$), the interaction is strong enough to correlate all layers,
as illustrated in Fig. \ref{fig:6}(a). On the other hand, for $d>2.5$,
each layer can transition independently between positive and negative
states, as shown in Fig. \ref{fig:6}(b). In this case, it is not
reasonable to calculate bulk magnetizations. 

In Fig. \ref{fig:7}(a), we present the critical temperature $T_{c}^{b}$
as a function of the distance between layers of thickness $d$ for
a bulk system with the magnetic interaction given by Eq. (\ref{eq:2}).
When, we compare these results with Fig. \ref{fig:4}, can observe
a rapid decay of $T_{c}$ for increasing values of $d$. It is worth
noting, and as showed in Fig. \ref{fig:6}, that Fig. \ref{fig:7}(a)
only presents data up to $d=2.0$, since from $d=2.5$ onwards the
layers are already disconnected and we no longer have the presence
of critical bulk behavior. Fig. \ref{fig:7}(b) illustrates the critical
temperature $T_{c}$ of the layers, demonstrating that $T_{c}^{l}$
is similar to $T_{c}^{b}$ for short values of $d$, and $T_{c}^{l}$
equals tends to a constant value of the 2D model, with the specific
$\alpha$, for high values of $d$.
\begin{center}
\begin{figure}
\begin{centering}
\includegraphics[scale=0.29]{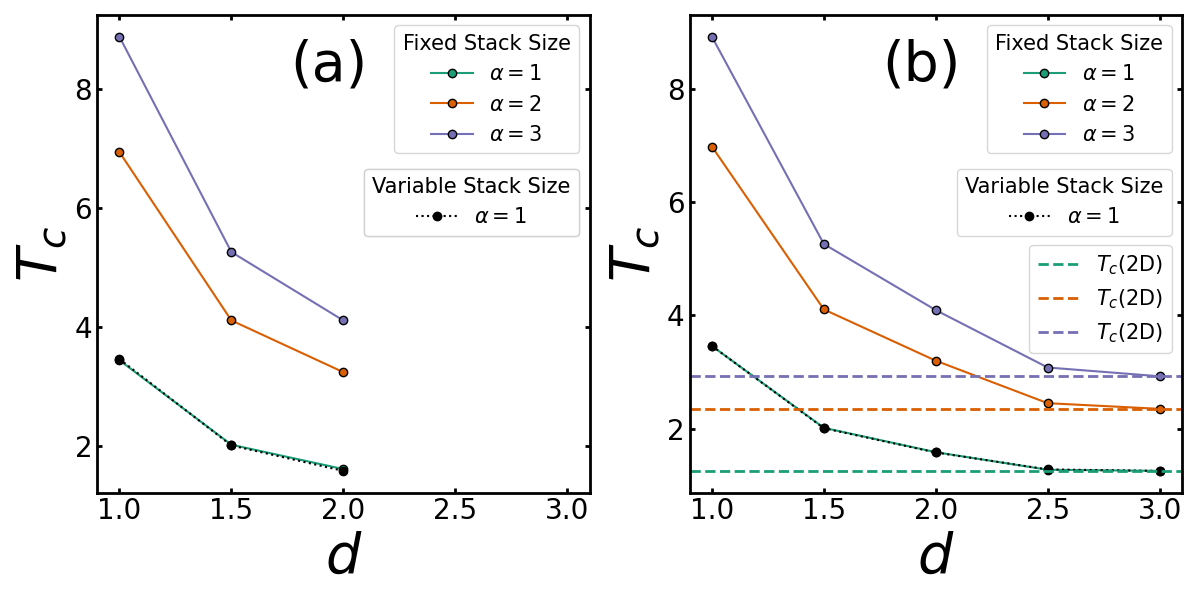}
\par\end{centering}
\caption{(a) Critical temperature multilayers $T_{c}^{b}$ versus interlayer
separations $d$ for three different values of the parameter $\alpha$,
as indicated in the figure. (b) Critical temperature of the layers
$T_{c}^{l}$ versus interlayer separations $d$ for three different
values of the parameter $\alpha$, as indicated in the figure. The
dashed lines represent the critical temperature $T_{c}(2D)$ of the
2D system for $d\gg3.0$. \protect\label{fig:7}}
\end{figure}
\par\end{center}

\begin{center}
\begin{figure*}
\begin{centering}
\includegraphics[scale=0.5]{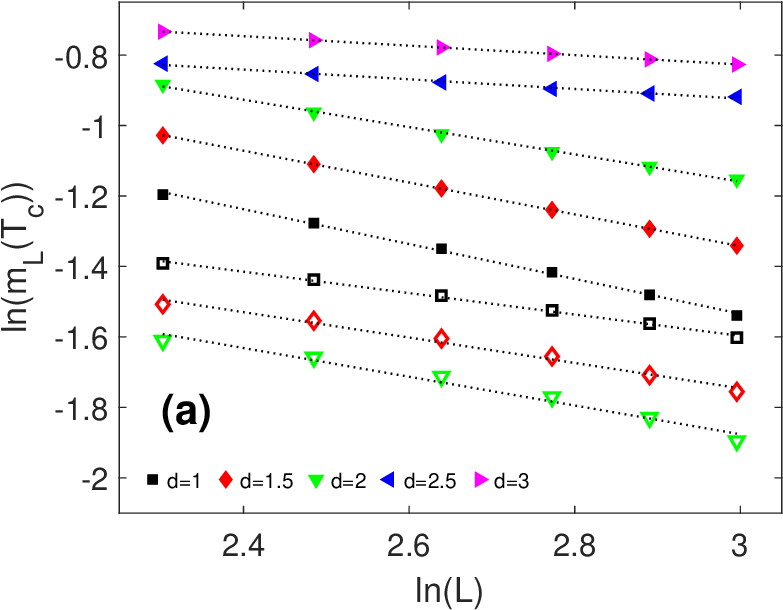}\hspace{0.1cm}\includegraphics[scale=0.5]{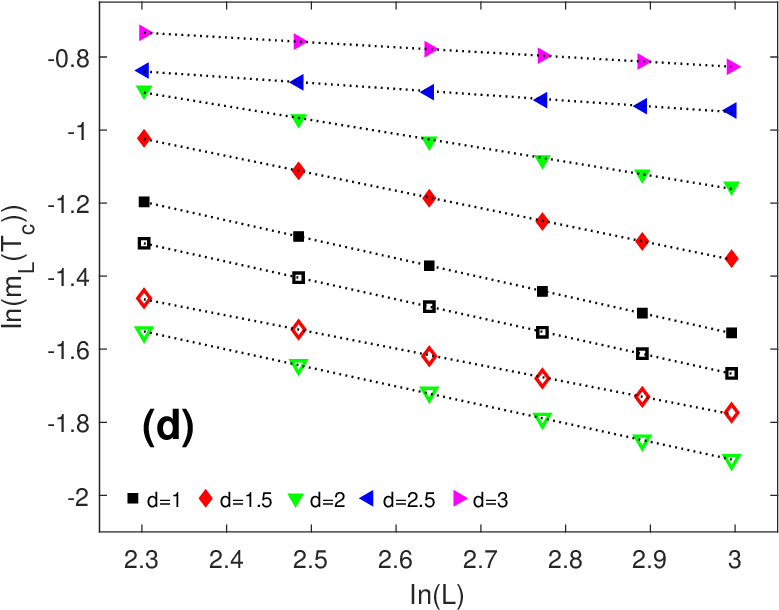}
\par\end{centering}
\begin{centering}
\vspace{0.1cm}
\par\end{centering}
\begin{centering}
\includegraphics[scale=0.5]{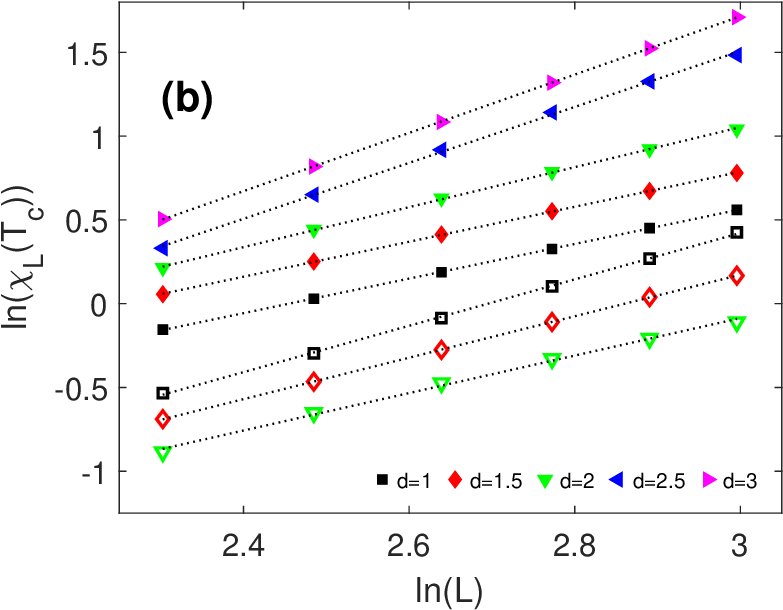}\hspace{0.1cm}\includegraphics[scale=0.5]{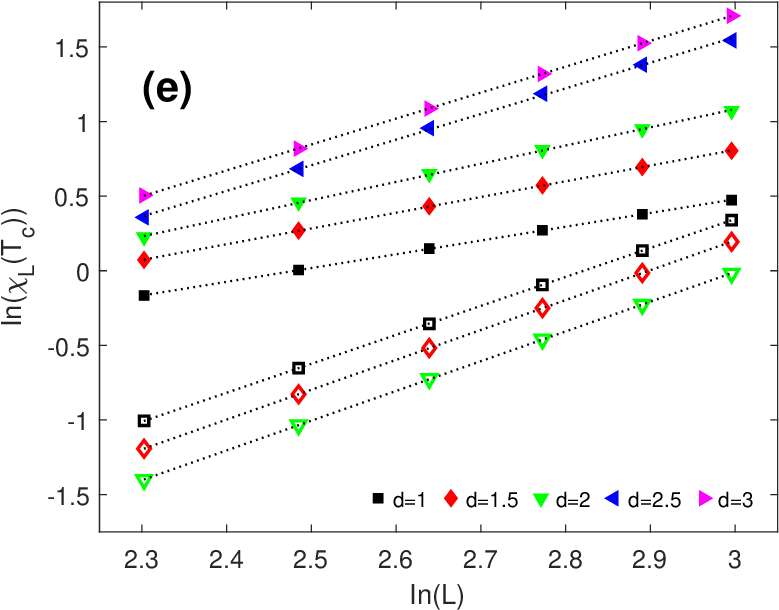}
\par\end{centering}
\begin{centering}
\vspace{0.1cm}
\par\end{centering}
\begin{centering}
\includegraphics[scale=0.5]{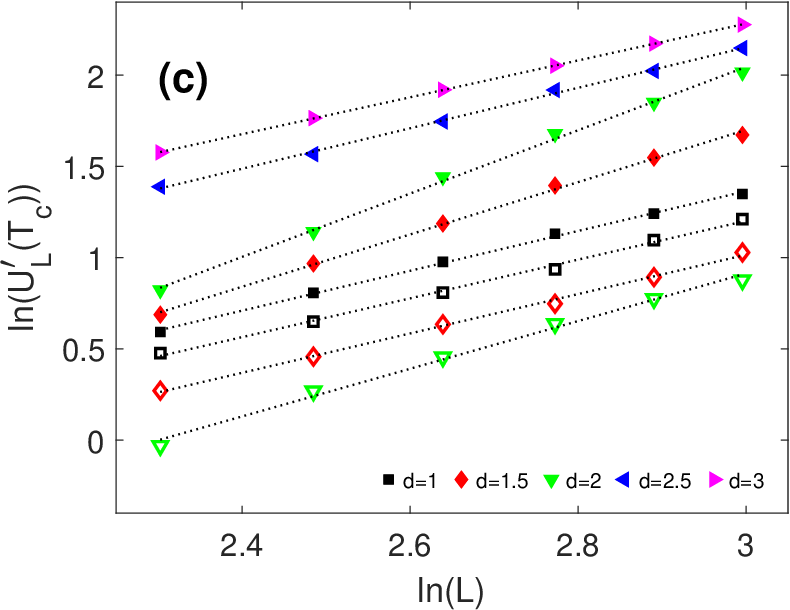}\hspace{0.1cm}\includegraphics[scale=0.5]{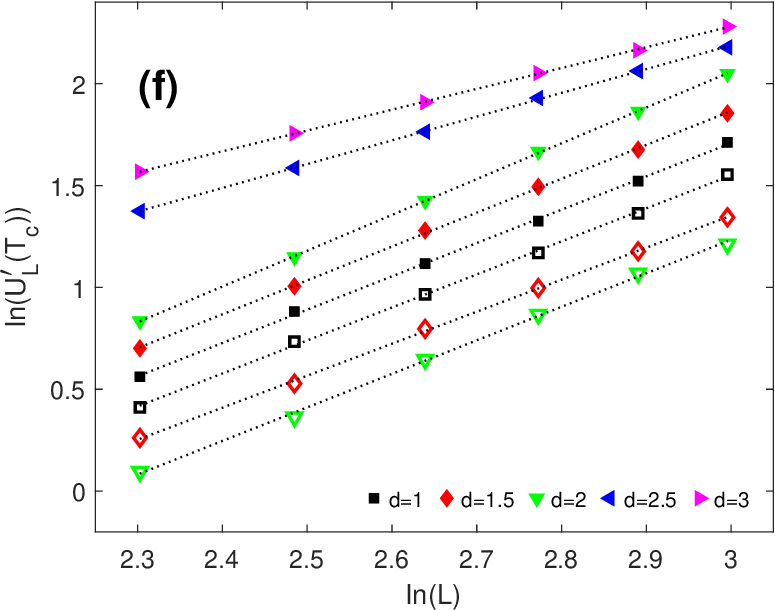}
\par\end{centering}
\caption{Log-log plots of $m_{L}^{\mu}(T_{c})$, $\chi_{L}^{\mu}(T_{c})$,
and $U_{L}^{\mu}(T_{c})$ versus $L$ at the critical point and for
different values of $d$, as presented in the figures. These results
were obtained for $\alpha=1.0$. Filled symbols represent the fits
for the layers, $\mu=l$, and empty symbols refers to the bulk system,
$\mu=b$. The dotted lines represent the best fit for the data points.
From these slopes, we have obtained the critical exponents $-\beta/\nu$,
$\gamma/\nu$, and $1/\nu$ as can be seen in Tabs. \ref{tab:1},
\ref{tab:2}, \ref{tab:3}, and \ref{tab:4}. The error bars are within
the symbol size. The panels (a), (b), and (c) correspond to the system
with $L_{z}=10$ $(N_{b}=10N_{l})$. The panels (d), (e), and (f)
correspond to the system with $L_{z}=L$ $(N_{b}=LN_{l})$. \protect\label{fig:8}}
\end{figure*}
\par\end{center}

From the MC simulations, we can also evaluate the critical exponents.
The finite-size scaling theory enables the extrapolation of data obtained
from simulations of finite systems to the thermodynamic limit. Since
one assumes that the thermodynamic quantities scale with the system
size $L$, close to $T_{c}$, as a power of $L$ multiplied by a nonsingular
function of the ratio between the critical exponents. Therefore, we
can calculate the critical exponents by analyzing the slope of the
best fit in the log-log plot using the scaling relations described
in Eqs. (\ref{eq:9}), (\ref{eq:10}), and (\ref{eq:11}). By examining
the slope of the log-log plot for the magnetization $m_{L}^{\mu}$
at the critical point for different lattice sizes $L$, as indicated
in Eq. (\ref{eq:9}), we determined the ratio $-\beta/\nu$, as depicted
in Fig. \ref{fig:8}(a) for $L_{z}$ stacked. Similarly, also for
$L_{z}$ stacked, the slope of the log-log plot of Eq. (\ref{eq:10})
provides us the ratio $\gamma/\nu$, see Figs. \ref{fig:8}(b). Additionally,
for the $\nu$ exponent associated with the correlation length of
the system, we employed the derivative of the Binder cumulant of the
scaling relations described in Eq. (\ref{eq:11}), where its slope
in the log-log plot yields the relation $1/\nu$, as illustrated in
Figs. \ref{fig:8}(c) for $L_{z}$ stacked. In the same way, but for
$L_{z}=L$, we have the best linear fit resulting in $-\beta/\nu$
, $\gamma/\nu$ and $1/\nu$, presented in Figs. \ref{fig:8}(d),
\ref{fig:8}(e), and \ref{fig:8}(f), respectively.

Given our focus on the slope of the log-log plot, we adjusted the
linear coefficients of the lines to separate them, making it easier
for the reader to visualize the fits. In Tabs. \ref{tab:1} and \ref{tab:2},
we list the values of the critical exponents for different values
of $\alpha$ and $d$, with the number of layers $L_{z}=10$ fixed
($N=10L^{2}$). We also present in Tabs. \ref{tab:3} and \ref{tab:4}
the data for $L_{z}=L$, i.e., $N_{b}=L^{3}$ only for $\alpha=1.0$,
since not much difference was observed in the critical behavior of
the system for different values of $\alpha$ when analyzing the system
with fixed $L_{z}$.
\begin{center}
\begin{table}
\caption{Critical exponents $\beta/\nu$, $\gamma/\nu$, and $1/\nu$ for different
values of $\alpha$, $d$, with $L_{z}=10$ $(N_{b}=10N_{l})$.{\footnotesize\protect\label{tab:1}}}

\centering{}{\small{}%
\begin{tabular}{cccccc}
\hline 
{\small$\alpha$} & {\small$d$} & {\small$T_{c}$} & {\small$\beta/\nu$} & {\small$\gamma/\nu$} & {\small$1/\nu$}\tabularnewline
\hline 
{\small 1.0} & {\small 1.0} & {\small$3.460\pm0.010$} & {\small$0.490\pm0.040$} & {\small$1.03\pm0.02$} & {\small$1.09\pm0.09$}\tabularnewline
{\small 1.0} & {\small 1.5} & {\small$2.015\pm0.005$} & {\small$0.450\pm0.004$} & {\small$1.04\pm0.02$} & {\small$1.43\pm0.15$}\tabularnewline
{\small 1.0} & {\small 2.0} & {\small$1.585\pm0.010$} & {\small$0.390\pm0.030$} & {\small$1.19\pm0.05$} & {\small$1.73\pm0.17$}\tabularnewline
{\small 1.0} & {\small 2.5} & {\small$1.275\pm0.010$} & {\small$0.140\pm0.030$} & {\small$1.67\pm0.10$} & {\small$1.11\pm0.15$}\tabularnewline
{\small 1.0} & {\small 3.0} & {\small$1.255\pm0.005$} & {\small$0.134\pm0.002$} & {\small$1.74\pm0.02$} & {\small$1.01\pm0.02$}\tabularnewline
\hline 
{\small 2.0} & {\small 1.0} & {\small$6.980\pm0.010$} & {\small$0.480\pm0.040$} & {\small$1.02\pm0.02$} & {\small$1.08\pm0.03$}\tabularnewline
{\small 2.0} & {\small 1.5} & {\small$4.105\pm0.005$} & {\small$0.500\pm0.020$} & {\small$1.00\pm0.02$} & {\small$1.35\pm0.13$}\tabularnewline
{\small 2.0} & {\small 2.0} & {\small$3.200\pm0.010$} & {\small$0.420\pm0.020$} & {\small$1.11\pm0.03$} & {\small$1.63\pm0.20$}\tabularnewline
{\small 2.0} & {\small 2.5} & {\small$2.450\pm0.020$} & {\small$0.190\pm0.060$} & {\small$1.64\pm0.15$} & {\small$1.39\pm0.08$}\tabularnewline
{\small 2.0} & {\small 3.0} & {\small$2.350\pm0.005$} & {\small$0.137\pm0.003$} & {\small$1.72\pm0.03$} & {\small$0.99\pm0.05$}\tabularnewline
\hline 
{\small 3.0} & {\small 1.0} & {\small$8.920\pm0.020$} & {\small$0.460\pm0.040$} & {\small$1.03\pm0.02$} & {\small$1.08\pm0.14$}\tabularnewline
{\small 3.0} & {\small 1.5} & {\small$5.260\pm0.010$} & {\small$0.480\pm0.020$} & {\small$1.00\pm0.01$} & {\small$1.32\pm0.13$}\tabularnewline
{\small 3.0} & {\small 2.0} & {\small$4.093\pm0.003$} & {\small$0.410\pm0.020$} & {\small$1.10\pm0.03$} & {\small$1.60\pm0.26$}\tabularnewline
{\small 3.0} & {\small 2.5} & {\small$3.080\pm0.010$} & {\small$0.210\pm0.060$} & {\small$1.61\pm0.13$} & {\small$1.47\pm0.07$}\tabularnewline
{\small 3.0} & {\small 3.0} & {\small$2.925\pm0.005$} & {\small$0.142\pm0.004$} & {\small$1.72\pm0.03$} & {\small$0.99\pm0.05$}\tabularnewline
\hline 
\end{tabular}}{\small\par}
\end{table}
\par\end{center}

\begin{center}
\begin{table}
\caption{Critical exponents $\beta/\nu$, $\gamma/\nu$, and $1/\nu$ for different
values of $\alpha$, $d$, and $N_{b}=LN_{l}$.{\footnotesize{} \protect\label{tab:2}}}

\centering{}%
\begin{tabular}{cccccc}
\hline 
{\small$\alpha$} & {\small$d$} & {\small$T_{c}$} & {\small$\beta/\nu$} & {\small$\gamma/\nu$} & {\small$1/\nu$}\tabularnewline
\hline 
{\small 1.0} & {\small 1.0} & {\small$3.44\pm0.02$} & {\small$0.30\pm0.04$} & {\small$1.38\text{\ensuremath{\pm0.06}}$} & {\small$1.06\pm0.15$}\tabularnewline
{\small 1.0} & {\small 1.5} & {\small$2.02\pm0.01$} & {\small$0.36\pm0.07$} & {\small$1.24\pm0.02$} & {\small$1.08\pm0.16$}\tabularnewline
{\small 1.0} & {\small 2.0} & {\small$1.61\pm0.01$} & {\small$0.41\pm0.13$} & {\small$1.12\pm0.12$} & {\small$1.31\pm0.23$}\tabularnewline
\hline 
{\small 2.0} & {\small 1.0} & {\small$6.94\pm0.01$} & {\small$0.29\pm0.03$} & {\small$1.36\pm0.05$} & {\small$1.00\pm0.17$}\tabularnewline
{\small 2.0} & {\small 1.5} & {\small$4.11\pm0.02$} & {\small$0.37\pm0.07$} & {\small$1.25\pm0.02$} & {\small$1.07\pm0.11$}\tabularnewline
{\small 2.0} & {\small 2.0} & {\small$3.24\pm0.02$} & {\small$0.45\pm0.11$} & {\small$1.10\pm0.08$} & {\small$1.11\pm0.19$}\tabularnewline
\hline 
{\small 3.0} & {\small 1.0} & {\small$8.88\pm0.03$} & {\small$0.30\pm0.03$} & {\small$1.35\pm0.04$} & {\small$1.08\pm0.13$}\tabularnewline
{\small 3.0} & {\small 1.5} & {\small$5.26\pm0.02$} & {\small$0.36\pm0.06$} & {\small$1.25\pm0.02$} & {\small$1.02\pm0.12$}\tabularnewline
{\small 3.0} & {\small 2.0} & {\small$4.11\pm0.02$} & {\small$0.28\pm0.09$} & {\small$1.23\pm0.02$} & {\small$1.10\pm0.18$}\tabularnewline
\hline 
\end{tabular}
\end{table}
\par\end{center}

\begin{center}
\begin{table}
\caption{Critical exponents $\beta/\nu$, $\gamma/\nu$, and $1/\nu$ for $\alpha=1.0$
and differents $d$, with $L_{z}=10$ $(N_{b}=10N_{l})${\footnotesize .\protect\label{tab:3}}}

\centering{}%
\begin{tabular}{ccccc}
\hline 
{\small$d$} & {\small$T_{c}$} & {\small$-\beta/\nu$} & {\small$\gamma/\nu$} & {\small$1/\nu$}\tabularnewline
\hline 
{\small 1.0} & {\small$3.460\pm0.010$} & {\small$0.518\pm0.007$} & {\small$0.957\pm0.006$} & {\small$1.64\pm0.11$}\tabularnewline
{\small 1.5} & {\small$2.008\pm0.002$} & {\small$0.476\pm0.020$} & {\small$1.05\pm0.01$} & {\small$1.66\pm0.08$}\tabularnewline
{\small 2.0} & {\small$1.580\pm0.010$} & {\small$0.38\pm0.04$} & {\small$1.22\pm0.04$} & {\small$1.76\pm0.05$}\tabularnewline
{\small 2.5} & {\small$1.280\pm0.005$} & {\small$0.16\pm0.03$} & {\small$1.72\pm0.09$} & {\small$1.17\pm0.05$}\tabularnewline
{\small 3.0} & {\small$1.255\pm0.005$} & {\small$0.134\pm0.001$} & {\small$1.74\pm0.01$} & {\small$1.02\pm0.01$}\tabularnewline
\hline 
\end{tabular}
\end{table}
\par\end{center}

\begin{center}
\begin{table}
\caption{Critical exponents $\beta/\nu$, $\gamma/\nu$, and $1/\nu$ for $\alpha=1.0$
and differents $d$ $(N_{b}=LN_{l})${\footnotesize .\protect\label{tab:4}}}

\centering{}%
\begin{tabular}{ccccc}
\hline 
{\small$d$} & {\small$T_{c}$} & {\small$-\beta/\nu$} & {\small$\gamma/\nu$} & {\small$1/\nu$}\tabularnewline
\hline 
{\small 1.0} & {\small$3.46\pm0.01$} & {\small$0.514\pm0.009$} & {\small$1.941\text{\ensuremath{\pm0.01}}$} & {\small$1.62\pm0.13$}\tabularnewline
{\small 1.5} & {\small$2.012\pm0.002$} & {\small$0.45\pm0.03$} & {\small$2.00\pm0.01$} & {\small$1.57\pm0.09$}\tabularnewline
{\small 2.0} & {\small$1.59\pm0.01$} & {\small$0.51\pm0.02$} & {\small$1.99\pm0.02$} & {\small$1.65\pm0.18$}\tabularnewline
\hline 
\end{tabular}
\end{table}
\par\end{center}

\begin{center}
\begin{figure*}
\begin{centering}
\includegraphics[scale=0.5]{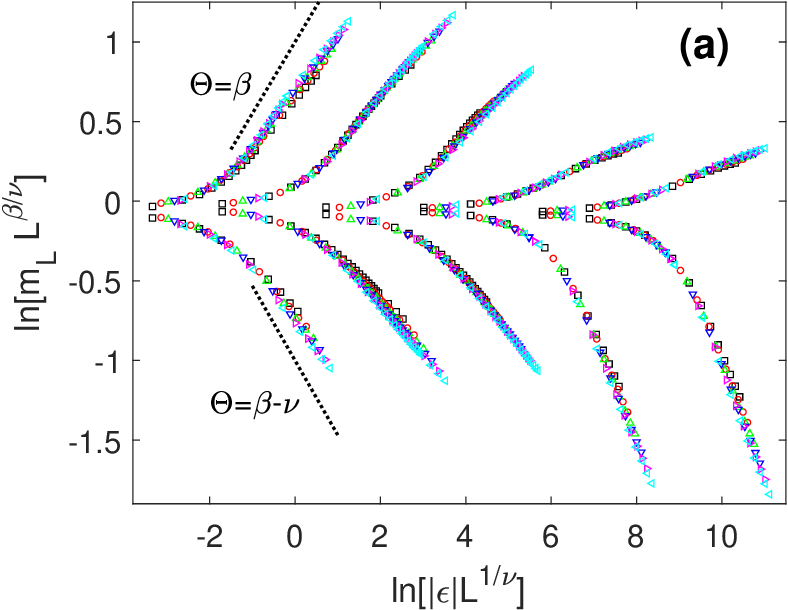}\hspace{0.1cm}\includegraphics[scale=0.5]{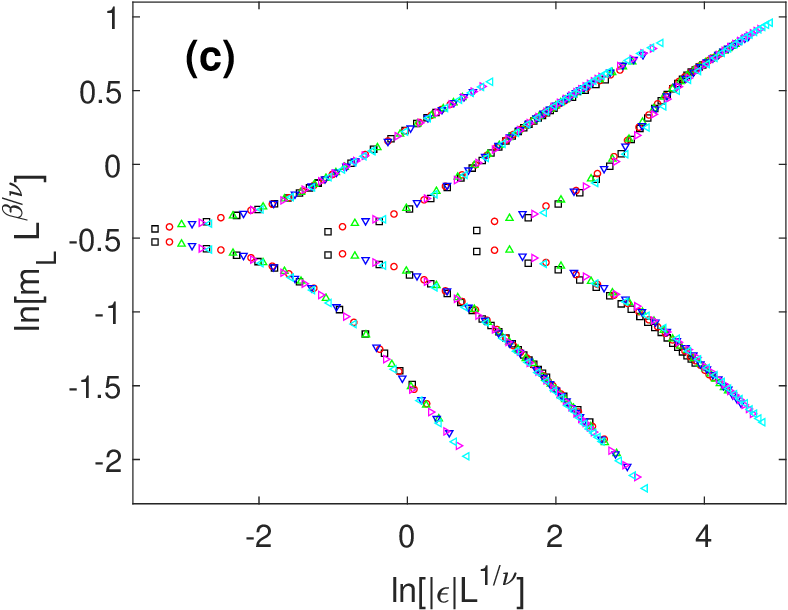}
\par\end{centering}
\begin{centering}
\vspace{0.1cm}
\par\end{centering}
\begin{centering}
\includegraphics[scale=0.5]{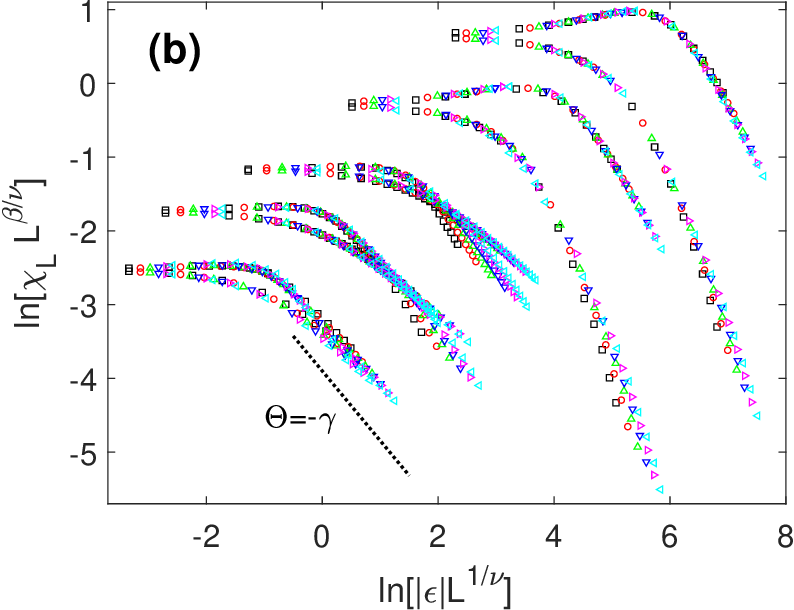}\hspace{0.1cm}\includegraphics[scale=0.5]{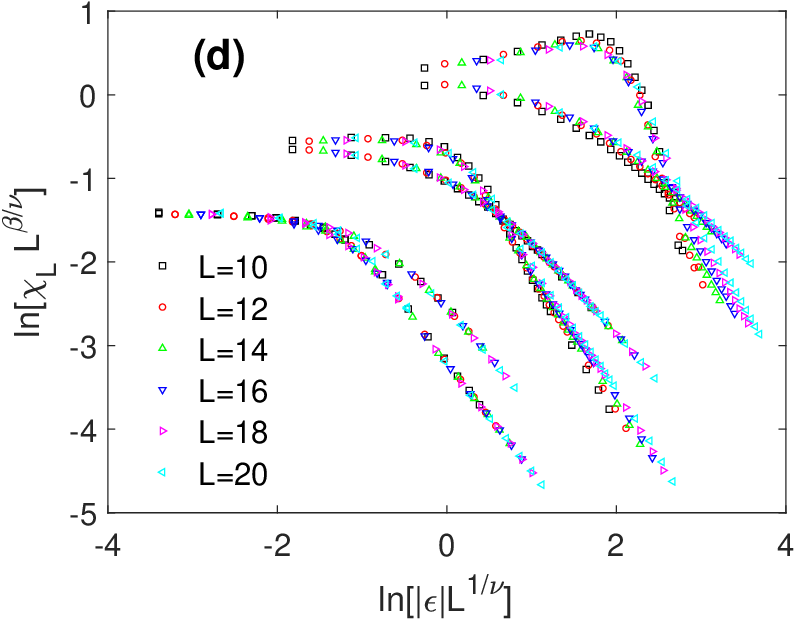}
\par\end{centering}
\caption{Data collapse near the critical point of the rescaled magnetization
$m_{L}^{b}L^{-\beta/\nu}$ and susceptibility $\chi_{L}^{b}L^{\gamma/\nu}$
as a function of $\left|\epsilon\right|L^{1/\nu}$ for various $d$
and $L$, as shown in panel (d). The panels (a) and (b) correspond
to layers while those panels (c) and (d) correspond to the bulk. Additionally,
in panels (a) and (c) the collapse sequence is from left to right
for $d=1.0$, $1.5$, $2.0$, $2.5$, $3.0$ and $d=1.0$, $1.5$,
$2.0$, respectively. In panels (b) and (d) the collapse sequence
is from bottom to top for $d=1.0$, $1.5$, $2.0$, $2.5$, $3.0$
and $d=1.0$, $1.5$, $2.0$, respectively. The log-log plots were
used to obtain the slope $\Theta$ of the asymptotic behavior of the
scaling functions, where the straight-dashed lines represent this
asymptotic behavior, as described by Eq. (\ref{fig:9}) and Eq. (\ref{eq:10}).
Here, we used $\alpha=1.0$ and $L_{z}=10$ ($N_{b}=10L^{2}$). \protect\label{fig:9}}
\end{figure*}
\par\end{center}

We also conducted another analysis aimed at determining the universal
functions of the magnetization Eq. (\ref{eq:9}) and the susceptibility
Eq. (\ref{eq:10}). The data collapse technique establishes scaling
and extracting associated critical exponents to equilibrium or non-equilibrium
phase transitions in many systems. These scaling functions represent
a data collapse \citep{23} for a specific value of $d$ and $\alpha$,
and various lattice sizes $L$. The critical exponents that best adjust
the curves correspond to the possible critical exponents.\textcolor{red}{{}
}For the values of $T$, we observe $\epsilon>0$ (ferromagnetic phase)
and $\epsilon<0$ (paramagnetic phase), resulting in two branches
in the data collapse, where the best collapse should occur near the
critical point. 
\begin{center}
\begin{figure*}
\begin{centering}
\includegraphics[scale=0.5]{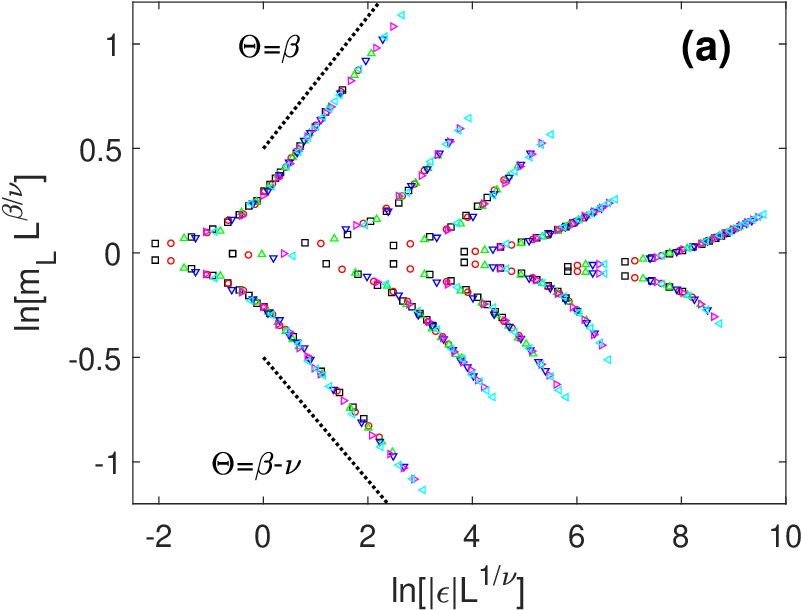}\hspace{0.1cm}\includegraphics[scale=0.5]{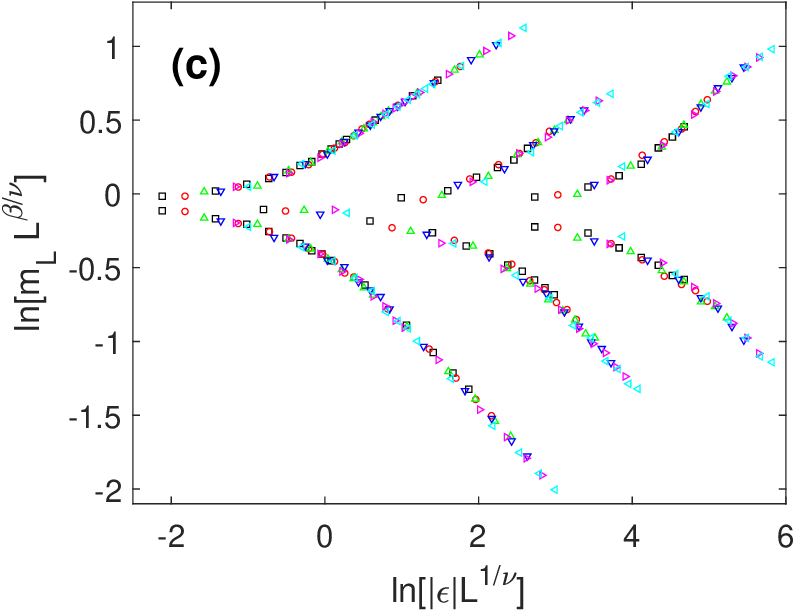}
\par\end{centering}
\begin{centering}
\vspace{0.1cm}
\par\end{centering}
\begin{centering}
\includegraphics[scale=0.5]{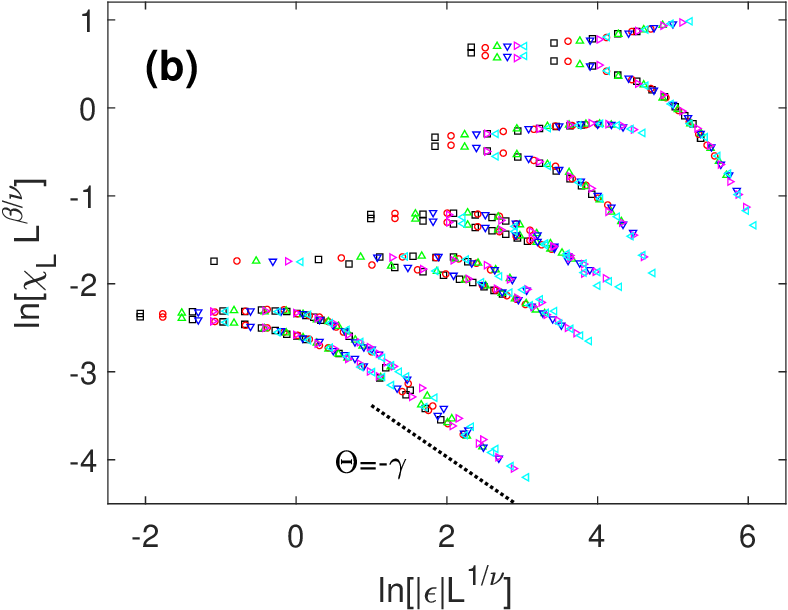}\hspace{0.1cm}\includegraphics[scale=0.5]{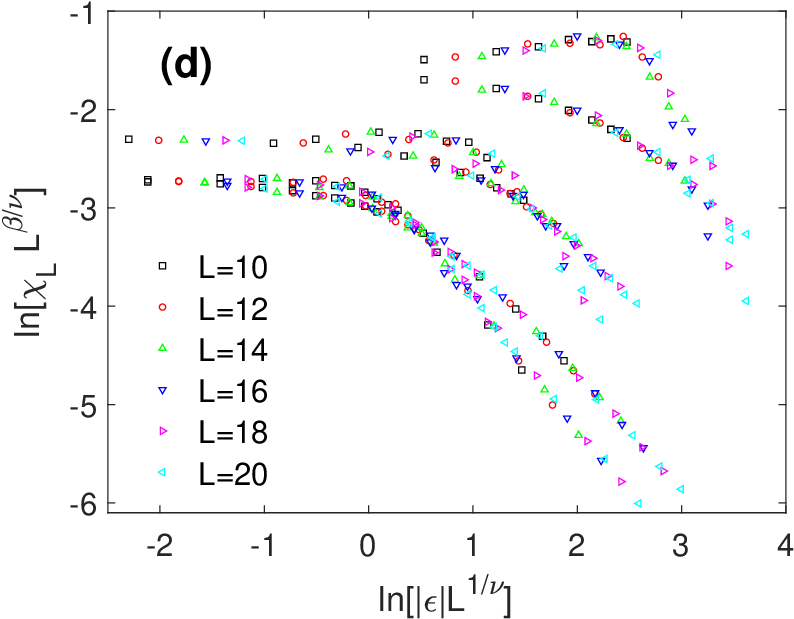}
\par\end{centering}
\caption{Data collapse near the critical point of the rescaled magnetization
$m_{L}^{\mu}L^{-\beta/\nu}$ and susceptibility $\chi_{L}^{\mu}L^{\gamma/\nu}$
as a function of $\left|\epsilon\right|L^{1/\nu}$ for various $d$
and $L$, as shown in panel (d). The panels (a) and (b) correspond
to layers while those panels (c) and (d) correspond to the bulk. Additionally,
in panels (a) and (c) the collapse sequence is from left to right
for $d=1.0$, $1.5$, $2.0$, $2.5$, $3.0$ and $d=1.0$, $1.5$,
$2.0$, respectively. In panels (b) and (d) the collapse sequence
is from bottom to top for $d=1.0$, $1.5$, $2.0$, $2.5$, $3.0$
and $d=1.0$, $1.5$, $2.0$, respectively. The log-log plots were
used to obtain the slope $\Theta$ of the asymptotic behavior of the
scaling functions, where the straight-dashed lines represent this
asymptotic behavior, as described by Eq. (\ref{fig:9}) and Eq. (\ref{eq:10}).
Here, we used $\alpha=1.0$ and $L_{z}=L$ ($N_{b}=L^{3}$).\protect\label{fig:10}}
\end{figure*}
\par\end{center}

In this way, the data collapse was used here to verify whether the
critical exponents, obtained from the linear fitting of the curves
near the critical point as a function of the system's linear size
in the log-log plot (see Fig. \ref{fig:8}), are compatible with the
system and describe the universality class of the system for the studied
values of $\alpha$ and $d$. For this purpose, in Fig. \ref{fig:9}
we present the collapsed curves using the exponents from Tables \ref{tab:1}
and \ref{tab:2}, that is, for the system with fixed $L_{z}=10$.
Fig. \ref{fig:9} (a) shows the magnetization curves related to the
average layer value, $m_{L}^{l}$, while Fig. \ref{fig:9} (b) shows
the collapse of the susceptibility of this magnetization, $\chi_{L}^{l}$.
Similarly, but now for the thermodynamic quantities related to the
bulk, in Fig. \ref{fig:9} (c) and (d) we present $m_{L}^{b}$ and
$\chi_{L}^{b}$ curves, respectively.

The verification of the exponents obtained from the linear fitting
through data collapse was also performed for the case where $L_{z}=L$,
i.e., the cubic system case. This is shown in Fig. \ref{fig:10},
where panel (a) presents the curves of $m_{L}^{l}$, and panel (b)
shows the curves of $\chi_{L}^{l}$, both related to the layers. In
Fig. \ref{fig:10} (c) and (d), for the bulk of the system, we present
the curves of $m_{L}^{b}$ and $\chi_{L}^{b}$, respectively. In the
analysis of the data collapse, we can see that the obtained exponents
are consistent with the critical exponents, as both Fig. \ref{fig:9}
and Fig. \ref{fig:10} provide good estimates of the scaling function
near the critical point, since the curves for various system sizes
merge into a single one in this regime.
\begin{center}
\begin{figure}
\begin{centering}
\includegraphics[scale=0.5]{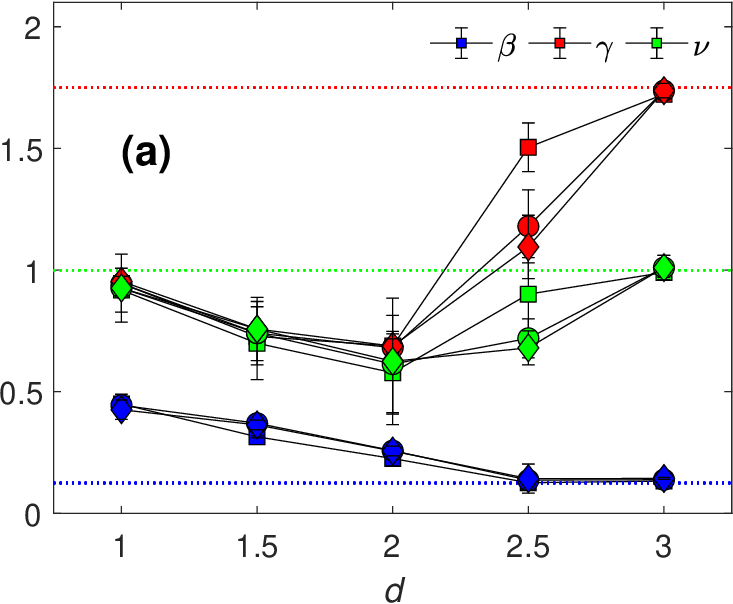}\hspace{0.1cm}\includegraphics[scale=0.5]{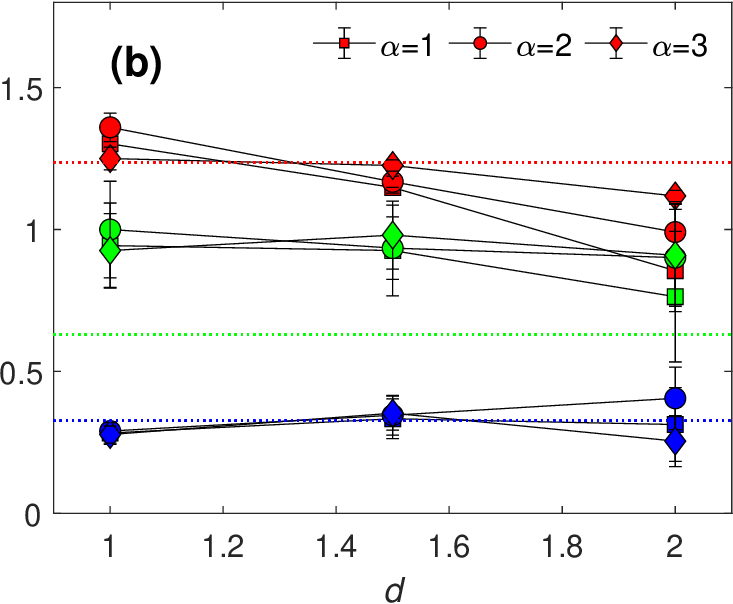}
\par\end{centering}
\caption{Critical exponents $\beta$, $\gamma$, and $\nu$ as a function of
the interlayer distance $d$, and some values of $\alpha$, as indicated
in panel (b). The dashed red, green, and blue lines correspond, respectively,
to the critical coefficients $\gamma$, $\nu$, and $\beta$ for the
Ising model in 2D (a), and in 3D (b). Here, $L_{z}=10$ and $N_{b}=10L^{2}$.
\protect\label{fig:11}}
\end{figure}
\par\end{center}

\begin{center}
\begin{figure}
\begin{centering}
\includegraphics[scale=0.5]{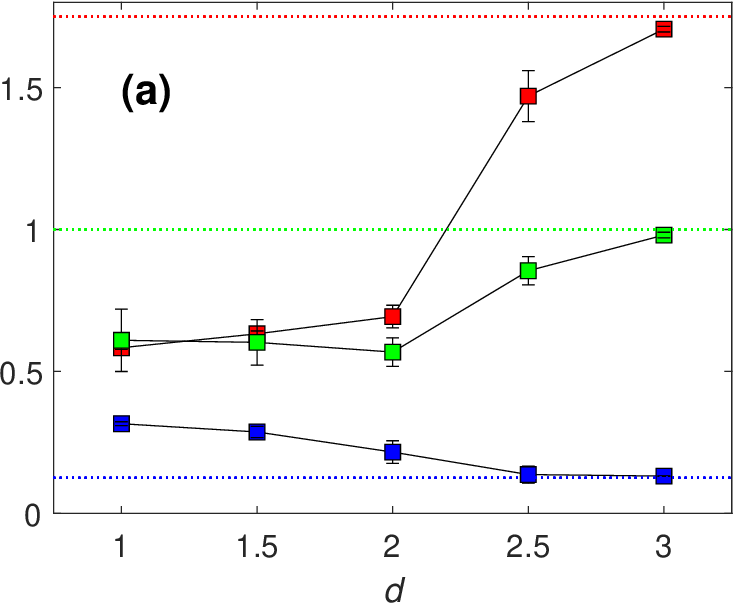}\hspace{0.1cm}\includegraphics[scale=0.5]{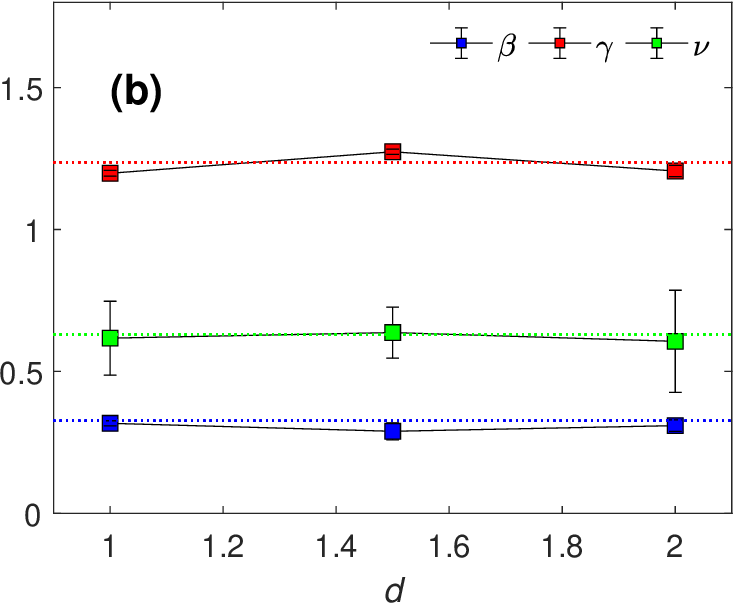}
\par\end{centering}
\caption{Critical exponents $\beta$, $\gamma$, and $\nu$ as a function of
the interlayer distance $d$. The dashed red, green, and blue lines
correspond, respectively, to the critical exponents $\gamma$, $\nu$,
and $\beta$ for the Ising model in 2D (a), and in 3D (b). Here, $\alpha=1.0$,
$L_{z}=L$, and $N_{b}=L^{3}$. \protect\label{fig:12}}
\end{figure}
\par\end{center}

We summarize the optimal values of the critical exponents, for the
case with fixed $L_{z}=10$, in a plot presented in Fig. \ref{fig:11}.
In Fig. \ref{fig:11}(a) we display the critical exponents for different
values of $\alpha$ and as a function of the interlayer distance $d$,
considering the simulations based on the average values of individual
layers. Fig. \ref{fig:11}(b) corresponds to the bulk system. We can
note from Fig. \ref{fig:11}(a) that for almost all values of interlayer
distance $d$ (except $d=2.5$), the different values of the parameter
$\alpha$ share the same critical exponents, especially when $d$
is greater than the imposed cutoff at $2.5$. At this point, the layers
become independent, and $\alpha$, $\beta$, and $\gamma$ converge
to the 2D Ising exponents. When the layers are closer and the interactions
more intense, significant changes occur in all critical exponents.
In Fig. \ref{fig:11}(b), we have the case of the bulk system with
fixed $L_{z}=10$, and even in the compact limit with $d=1.0$, although
$\beta$ converges to the 3D Ising value, the average value of $\gamma$
deviates slightly, and $\nu$ is far from the values identified in
the literature.

With these results, we can observe that in the case where the layers
are analyzed, for low values of $d$, the strong interaction between
the layers changes the universality of the 2D system, such that we
reach the critical dimension $D_{c}=4$ of the Ising model. This is
evident because at $d=1$, $\beta\approx0.5$, $\gamma\approx1.0$,
and $\nu\approx1.0$ are the exponents of the mean-field approximation.
However, as $d$ increases, the interactions between the layers weaken
and gradually modify the exponents until we reach the universality
of the 2D Ising model, as the layers become isolated. When analyzing
the bulk system, since we fixed $L_{z}=10$ but the layers keep changing
in size, we do not have a well-defined linear size for the system,
and the exponents are not fully compatible with the 3D Ising model.

Now, for the simulations performed with a variable $L_{z}=L$, the
optimal values of the critical exponents are presented in Fig \ref{fig:12}.
This can be seen in Fig. \ref{fig:12}(a) the critical exponents for
$\alpha=1.0$ as a function of the interlayer distance $d$, considering
simulations based on the average values of individual layers, while
Fig. \ref{fig:12}(b) corresponds to the bulk system. From Fig. \ref{fig:12}(a),
it is easy to see that the critical exponents obtained as averages
from individual layers when $L_{z}$ is variable show similar qualitative
behavior to that when $L_{z}$ is fixed (see Fig. \ref{fig:11}(a)),
even reaching the 2D critical exponents when layers become independent
at $d=3.0$, but not exhibiting mean-field critical behavior at $d=1.0$,
as the system now has a well-defined dimension, a cubic system. In
Fig. \ref{fig:12}(b), we can observe that when performing simulations
with varying $L_{z}$ and with periodic boundary conditions, the values
of the 3D critical exponents are reached in the compact limit at $d=1.0$
and the interaction between neighboring layers is intense enough to
maintain these even at greater distances ($d=2.0$).

\section{Conclusions\protect\label{sec:Conclusions}}

In this work, we have used MC simulations and the finite-size scaling
theory to explore the critical behaviors and thermodynamic properties
of a multilayer system. Our study specifically addressed how variations
in the interlayer distance influence the magnetic coupling and the
phase transitions in these systems. We conducted simulations with
fixed and variable numbers of layers $L_{z}$, analyzing the behavior
of the layers as well as the bulk system. For the fixed value of $L_{z}=10$,
the layers interact strongly in $1.0\leq d\leq2.0$, causing the exponents
to deviate from the theoretical values for the 2D Ising model. However,
in this case, since we do not have a well-defined dimension in the
system, we find exponents similar to those of the mean-field approximation
at $d=1.0$. In this regime, it is also observed that the exponents
do not have a strong dependence on $\alpha$. As $d$ increases and
we reach the cutoff value ($d=r_{ij}=2.5$), the exponents become
more sensitive to the values of $\alpha$. Nevertheless, when $d>2.5$,
the layers become independent, and the critical exponents with different
$\alpha$ values converge to those of the 2D Ising model. On the other
hand, when analyzing the bulk system, the difference between the number
of layers $L_{z}$ and the layer sizes $N=L\times L$, prevents the
system from exhibiting the universality class of the 3D Ising model,
especially regarding the exponent related to the correlation length.

On the other hand, when varying $L_{z}$, so that we are now dealing
with a cubic system, we have a well-defined dimension in the system.
The critical exponent related to the correlation length, both in the
case of the layers and in the bulk, converges to the value of the
3D Ising model in the range $1.0\leq d\leq2.0$. This change in the
exponent $\nu$ prevents us from obtaining critical exponents resembling
those of the mean-field approximation at $d=1.0$ when analyzing only
the layers. However, for $d>2.5$, we still find exponents of the
2D Ising model, as in this regime the layers are isolated. Nevertheless,
because both the number of layers and the layer sizes increase uniformly,
in the bulk system analysis, we always find critical exponents of
the 3D Ising model.

Finally, based on the data from Tables \ref{tab:1} and \ref{tab:3},
we can observe that the critical behavior of the layers system, whether
with a fixed or varying $L_{z}$, shares a weak universality class,
as the ratios of the exponents $\beta/\nu$ and $\gamma/\nu$ are
equivalent, within the respective margins of error. However, the exponent
$\nu$ depends on the behavior of $L_{z}$. On the other hand, the
critical behavior of the bulk system, as seen in Tables \ref{tab:2}
and \ref{tab:4}, is significantly affected by whether $L_{z}$ is
fixed or varies with the layer sizes. We also observe a weak universality
class relating both the layers and the bulk system, as in the regime
$1.0\leq d\leq2.0$, there is a strong interaction between the layers,
and both systems share the same value for the exponent $\nu$, within
the appropriate margin of error.
\begin{acknowledgments}
This work was financially supported by the Fundação Coordenação de
Aperfeiçoamento de Pessoal de Nível Superior (CAPES), and Conselho
Nacional de Desenvolvimento Científico e Tecnológico (CNPq) of Brazil
(Process No. 140141/2024-3).
\end{acknowledgments}

\end{document}